\documentclass[prd,aps,a4paper,nofootinbib,twocolumn]{revtex4}  
\newif\ifusesec
\usesectrue  
   
\usepackage{graphicx} 
\usepackage{mathrsfs}
\usepackage{amsmath,amsfonts,amssymb}

\newcommand{\beq}{\begin{equation}}
\newcommand{\eeq}{\end{equation}}
\newcommand{\bea}{\begin{eqnarray}}
\newcommand{\eea}{\end{eqnarray}}

\begin{document}

\title{Spin-orbit contribution to radiative losses for spinning binaries with aligned spins}

\author{Donato Bini$^{1,2}$, Andrea Geralico$^1$, Piero Rettegno$^3$}
  \affiliation{
$^1$Istituto per le Applicazioni del Calcolo ``M. Picone,'' CNR, I-00185 Rome, Italy\\
$^2$INFN, Sezione di Roma Tre, I-00146 Rome, Italy\\
$^3$INFN Sezione di Torino, Via P. Giuria 1, 10125 Torino, Italy
}

\date{\today}

\begin{abstract}
We compute the leading order contribution to radiative losses in the case of spinning binaries with aligned spins due to their spin-orbit interaction. 
The orbital average along hyperboliclike orbits is taken through an appropriate spin-orbit modification to the quasi-Keplerian parametrization for nonspinning bodies, which maintains the same functional form, but with spin-dependent orbital elements.
We perform consistency checks with existing PN-based and PM-based results. In the former case, we compare our expressions for both radiated energy and angular momentum with those obtained in [JHEP \textbf{04}, 154 (2022)] by applying the boundary-to-bound correspondence to known results for ellipticlike orbits, finding agreement.
The linear momentum loss is instead newly computed here.
In the latter case, we also find agreement with the low-velocity limit of recent calculations of the total radiated energy, angular momentum and linear momentum in the framework of an extension of the worldline quantum field theory approach to the classical scattering of spinning bodies at the leading post-Minkowskian order [Phys. Rev. Lett. \textbf{128}, no.1, 011101 (2022), Phys. Rev. D \textbf{106}, no.4, 044013 (2022)].
We get exact expressions of the radiative losses in terms of the orbital elements, even if they are at the leading post-Newtonian order, so that their expansion for large values of the eccentricity parameter (or equivalently of the impact parameter) provides higher-order terms in the corresponding post-Minkowskian expansion, which can be useful for future crosschecks of other approaches.
\end{abstract}

\maketitle

\section{Introduction}

The importance of including spin effects in the description of the dynamics of a two-body system has been known for a long time. In the pioneering work of Kidder \cite{Kidder:1995zr} it was shown that the effects of spin-orbit and spin-spin interactions on the inspiral of a coalescing binary system of spinning compact objects may be relevant for rapidly rotating bodies, since they induce post-Newtonian (PN) corrections which are comparable with those of the nonspinning counterparts.

The presence of nonnegligible spins can strongly affect the binary dynamics.
In particular, spin vectors not aligned with the orbital angular momentum cause the orbital plane to precess, leading to modulations in the observed shape of the waveform (see also Ref. \cite{Apostolatos:1994mx}).
In addition, the spins of the bodies modify the radiative multipole moments, so that they directly contribute to the amplitude and accumulated phase of the waveform as well as to the losses of energy, angular momentum and linear momentum from the system.

In the present paper we focus on spin-orbit effects, which are linear in the spins of the binary.
The study of the spin-orbit coupling was already carried out at the leading PN order in Ref. \cite{Kidder:1995zr}, later extended to the next-to-leading order in Refs. \cite{Tagoshi:2000zg,Faye:2006gx,Damour:2007nc,Tessmer:2010hp,Porto:2010tr,Levi:2010zu} by using different approaches (including effective-field-theory-based \cite{Goldberger:2004jt,Porto:2005ac} and amplitude-based \cite{Bern:2020buy} methods), also taking into account radiation-reaction effects \cite{Will:2005sn,Blanchet:2006gy,Wang:2011bt,Maia:2017gxn}.
Higher-order extensions have been obtained too over the years \cite{Barausse:2011ys,Bohe:2012mr,Bohe:2013cla,Hartung:2013dza,Levi:2015uxa,Levi:2020kvb,Kim:2022pou,Mandal:2022nty}.

Most of the efforts have been devoted so far to the case of quasicircular motion, since the emission of gravitational radiation tends to circularize the orbits of the binary.
The spin-orbit contributions to the radiative losses have been computed first by Kidder at the leading PN order for general orbits, further specialized to circular motion.
These contributions turn out to be 1.5PN order relative to the leading Newtonian order in the case of the energy and angular momentum, and 0.5PN order in the case of the linear momentum.
Next-to-leading order corrections have been then obtained in Ref. \cite{Racine:2008kj} (see also Ref. \cite{Gopakumar:2011zz}). 
The state of the art for ellipticlike orbits is the next-to-next-to-leading order spin-orbit and spin-spin corrections to the energy loss at 4PN level, while 3PN for both angular momentum and linear momentum losses \cite{Bohe:2013cla,Cho:2021mqw,Cho:2021arx,Cho:2022syn}. 

We compute here the contributions to the orbital average of the energy, angular momentum and linear momentum fluxes along hyperboliclike orbits of a two-body system due to their leading-order spin-orbit interaction in the simplest case of spins with constant magnitude and orthogonal to the orbital plane.
This special situation allows for using a minimal modification to the quasi-Keplerian parametrization of the orbit for nonspinning bodies \cite{dd} (see Refs. \cite{Gergely:1998sr,Konigsdorffer:2005sc,Cho:2019brd} for the case of ellipticlike motion).
Our results are in agreement with the low-velocity limit of recent calculations of the total radiated energy, angular momentum and linear momentum obtained in Refs. \cite{Jakobsen:2021lvp,Riva:2022fru} by using an extension of the worldline quantum field theory approach to the classical scattering of spinning bodies at the leading post-Minkowskian order.
Although valid at the leading PN order our expressions for the radiative losses are exact functions of the orbital elements.
Therefore, they can be expanded in PM sense as series expansions in the large-eccentricity parameter (or equivalently large-impact parameter), which give higher-order PM terms useful for crosschecking companion amplitude-based and PM-based computations.

We will use a mostly positive signature of the metric $(-+++)$.
The two bodies have masses $m_1$ and $m_2$ (with $m_1>m_2$) and spin vectors ${\mathbf S}_1$ and ${\mathbf S}_2$. Standard mass ratio notations will be used for the total mass $M=m_1+m_2$, the reduced mass $\mu=m_1m_2/M$, the symmetric mass-ratio $\nu=\mu/M$, mass difference $\delta m=m_1-m_2$, as well as for the two symmetric combinations of spins
\beq
{\mathbf S}={\mathbf S}_1+{\mathbf S}_2\,,\qquad 
{\mathbf S}_*=\frac{m_2}{m_1}{\mathbf S}_1+\frac{m_1}{m_2}{\mathbf S}_2\,.
\eeq
We work in harmonic coordinates such that the orbital plane is the $x-y$ plane, and the direction of the spins (as well as orbital angular momentum) is the $z$-axis
\beq
\label{SSstar}
{\mathbf S}=Se_z\,,\qquad {\mathbf S}_*=S_*e_z\,.
\eeq
The spin vectors also have constant magnitude.
To simplify expressions we mostly work in units of $G$, $M$ and $c$, except for comparison with existing results.

\section{Spin-modified quasi-Keplerian parametrization of the orbit}

In the case of a spinning binary system the finite size of the bodies introduces an ambiguity in the definition of the center of mass of each body, corresponding to different choices of spin supplementary conditions (see, e.g., Appendix A of Ref. \cite{Kidder:1995zr}).
It is customary to assume the so called covariant supplementary conditions $S_A^{\mu\nu}u_{A\,\nu}=0$, where $u_{A\,\nu}$  (with $u_{A\,\mu}u_A^{\mu}=-1$) denotes the four velocity of the center-of-mass worldline of body $A$, and $S_A^{\mu\nu}$ the associated (antisymmetric) spin tensor.
To linear order in spin (which is the case we are interested in) the kinematical four momentum is aligned with the four velocity, $p_A=m_Ac\,u_A+O(s_A^2)$, and the spin magnitudes $s_A^2=\frac12S_A^{\mu\nu}S_{A\,\mu\nu}$ remain constant as a consequence of the spin evolution equations if higher-order multipole couplings are neglected.
The spin vector $S_A$ associated with the spin tensor is defined by spatial duality with respect to $u_A$, $S_A^\mu=\frac12u_{A\nu}\eta^{\nu\mu\alpha\beta}S_{A\alpha\beta}$, so that it is orthogonal to $u_A$, and has only spatial components in a frame adapted to $u_A$. One can then define a constant-magnitude 3-dimensional spin vector ${\mathbf S}_A^c$ which satisfies an ordinary precession equation $\frac{d{\mathbf S}_A^c}{dt}={\mathbf \Omega}_A\times {\mathbf S}_A^c$ \cite{Damour:2007nc,Bohe:2012mr}.
The angular velocity ${\mathbf \Omega}_A$ vanishes at the Newtonian order, i.e., starts at 1PN order, and is directly computable from the spacetime metric.
The transformation from the original spin variables to those with constant magnitude is also 1PN \cite{Faye:2006gx}, so that at the leading PN order we are working with they coincide.
We further assume that the spins are initially aligned with the orbital angular momentum, implying that $\frac{d{\mathbf S}_A}{dt}=0$ (see Eq. (2.4) of Ref. \cite{Kidder:1995zr}), so that their direction does not change during the evolution.

Up to 2.5PN order the motion is conservative, and the energy and total angular momentum are conserved.
The leading-order spin-orbit acceleration is 1.5PN beyond the Newtonian one. 
The equations of motion in harmonic coordinates in the center of mass frame are given by (see Eqs. (2.1) and (2.2) of Ref. \cite{Kidder:1995zr})
\beq
{\mathbf a} ={\mathbf a}_\mathrm{N} + \frac{1}{c^2}{\mathbf a}_\mathrm{1PN} + \frac{1}{c^3}{\mathbf a}_\mathrm{SO}\,,
\eeq
with
\bea
\label{accs}
{\mathbf a}_\mathrm{N} &=& -\frac{M}{r^2}{\mathbf n}
\,,\nonumber\\
{\mathbf a}_\mathrm{1PN} &=& -\frac{M}{r^2} \left(A^{\rm 1PN}{\mathbf n}+ B^{\rm 1PN}\dot r {\mathbf v} \right)
\,,\nonumber\\
{\mathbf a}_\mathrm{SO} &=& \frac{6}{r^3}\left[({\mathbf n} \times {\mathbf v})\cdot\left( {\mathbf S} + {\mathbf S}_*\right)\right]{\mathbf n}    - \frac{1}{r^3}{\mathbf v} \times \left(4{\mathbf S} + 3{\mathbf S}_*\right)\nonumber\\
&+& 
\frac{3\dot r}{r^3}{\mathbf n} \times \left(2{\mathbf S} +{\mathbf S}_*  \right)\,,
\eea
where ${\mathbf x} = r{\mathbf n}$ and ${\mathbf v}=\frac{d {\mathbf x}}{dt}$, with $r=|{\mathbf x}|$ and $\dot r={\mathbf v}\cdot {\mathbf n}$, and
\bea
A^{\rm 1PN}&=&(1+3\nu)v^2 - \frac{3}{2}\nu\dot r^2 - 2(2+\nu)\frac{M}{r}\,,  \nonumber\\ 
B^{\rm 1PN}&=&-2 (2-\nu)\,.
\eea
The spin-orbit acceleration then reduces to ${\mathbf a}_\mathrm{SO} ={\mathbf a}_\mathrm{SO}^x {\mathbf e}_x+{\mathbf a}_\mathrm{SO}^y {\mathbf e}_y$, with
\bea
{\mathbf a}_\mathrm{SO}^x&=& \frac{2}{r^3} \left[ r\cos \phi \dot \phi  \left(S + \frac32 S_*\right) +  \sin\phi \dot r S  \right]\,,\nonumber\\
{\mathbf a}_\mathrm{SO}^y&=&  \frac{2}{r^3} \left[  r\sin\phi \dot \phi \left(S + \frac32 S_*\right)  -  \cos\phi\dot r  S  \right]\,,
\eea
for constant spin vectors aligned with the $z$-axis, Eq. \eqref{SSstar}.

It is easy to show that in the special case considered here one can still have a quasi-Keplerian parametrization of the hyperboliclike orbit (which is currently known up to the 3PN order \cite{Cho:2018upo,Bini:2022enm}), i.e.,
\bea
\label{orbit}
\bar n t &=&e_t\sinh v-v\,,\nonumber\\ 
r&=&\bar a_r (e_r\cosh v-1)\,,\nonumber\\
\phi&=&2 K {\rm arctan}\left(\sqrt{\frac{e_\phi+1}{e_\phi-1}}\tanh \frac{v}{2} \right)\,,
\eea
where the orbital elements $\bar n$, $K$, $\bar a_r$, $e_t$, $e_r$ and $e_\phi$ have 1.5PN spin corrections.
Expressing the orbital elements in terms of $e_r$ and $\bar a_r$ we find
\bea
K&=&1+\frac{3}{\bar a_r (e_r^2-1)}\eta^2-\frac{4S+3S_*}{[\bar a_r (e_r^2-1)]^{3/2}}\eta^3\,, \nonumber\\
\bar n&=& \frac{1}{\bar a_r^{3/2}}-\frac{\nu-9}{2\bar a_r^{5/2}}\eta^2 -\frac32 \frac{2S+S_*}{\bar a_r^3 (e_r^2-1)^{1/2}}\eta^3\,,\nonumber\\
e_t&=& e_r-\frac{e_r\left(3\nu-8\right)}{2\bar a_r}\eta^2 -\frac{e_r(2S+S_*)}{\bar a_r^{3/2}(e_r^2-1)^{1/2}}\eta^3\,,\nonumber\\
e_\phi&=& e_r-\frac{\nu e_r}{2\bar a_r}\eta^2 +\frac{e_rS_*}{\bar a_r^{3/2}(e_r^2-1)^{1/2}}\eta^3 \,,
\eea
with $\eta=1/c$ a place-holder, and $S,S_*\sim \frac{S}{M^2},\frac{S_*}{M^2}$ are adimensionalized by using the total mass of the system. 
The corresponding relations for ellipticlike motion have been obtained in Ref. \cite{Konigsdorffer:2005sc} in Arnowitt, Deser, and Misner-type coordinates and for arbitrary spin orientations (see also Ref. \cite{Cho:2019brd}).
In a sense, at this order, the spin structure of the bodies manifests itself as an additional structure of the orbit, i.e. modifies the eccentricities too, besides (as expected) the  various gauge-invariant quantities as the periastron precession, the radial period, etc. This effect can be interpreted qualitatively in terms of the equivalence principle.

The conserved energy and total angular momentum are given by (see Eqs. (2.6)--(2.9) of Ref. \cite{Kidder:1995zr})
\beq
E=E_{\rm N}+\frac{1}{c^2}E_{\rm 1PN}+ \frac{1}{c^3}E_{\rm SO}\,,
\eeq
with
\begin{align}
	E_{\rm N} =& \mu \left(\frac{1}{2}v^2-\frac{M}{r}\right)\,, \nonumber \\
	E_{\rm 1PN} =& \mu \Bigg[ \frac{3}{8}\left(1-3\nu\right)v^4+\frac{1}{2}\left(3+\nu\right)\frac{v^2}{r} \nonumber \\
	&+ \frac{1}{2}\nu \frac{M}{r}\dot{r}^2+\frac{1}{2}\left(\frac{M}{r}\right)^2 \Bigg]\,, \nonumber \\
E_{\rm SO}=&\frac{L_{\rm N}}{r^3}S_*\,,
\end{align}
and 
\beq
{\mathbf J}={\mathbf L}+{\mathbf S}\,,
\eeq
where
\beq
{\mathbf L}={\mathbf L}_{\rm N}+\frac{1}{c^2}{\mathbf L}_{\rm 1PN}+ \frac{1}{c^3}{\mathbf L}_{\rm SO}\,,
\eeq
with ${\mathbf L}_{\rm N}=\mu\, {\mathbf x}\times {\mathbf v}$, and 
\begin{align}
	{\mathbf L}_{\rm 1PN} =& L_{\rm N} \left[ \frac{1}{2}\left(1-3\nu\right)v^2+\left(3+\nu\right)\frac{M}{r} \right]\,, \nonumber \\
	{\mathbf L}_{\rm SO}=&\frac{\mu}{M}\left[-\frac{M}{r}\left(2S+S_*\right)+\frac{1}{2} S_* v^2\right] \,,
\end{align}
respectively.
Since the spins are nonprecessing, the orbital angular momentum does not precess too, and is conserved as well.

The orbital elements can then be expressed in terms of the conserved energy and orbital angular momentum by using the relations 
\bea
\label{ar_e_r_vs_EL}
\bar{a}_r&=&\frac{1}{2 \bar E}
\left[1+\frac{1}{2}\eta^2\left(7-\nu\right)\bar E-\eta^3\frac{2
\bar E}{L}\left(2S+S_*\right)\right]
\,,\nonumber\\
e_r&=&\sqrt{1+2\bar E L^2}+\eta^2\frac{\bar{E}\left[-6 + \nu +
    \frac{5}{2}\left(-3+\nu\right)\bar{E}L^2\right]}{\sqrt{1+2\bar{E}L^2}}\nonumber\\
&+&
\eta^3\frac{2 \bar{E}\left[4 S + 2 S_* + \left(4S+S_*\right)\bar{E}L^2\right]}{L \sqrt{1+2\bar{E}L^2}}\,,
\eea
with $\bar E=(E-M)/\mu$ being the binding energy and $L$ the dimensionless orbital angular momentum (rescaled by $M\mu$).

\section{Spin-orbit corrections to the radiative losses}

According to the multipolar-post-Minkowskian (MPM) formalism the radiative fluxes of energy, linear momentum and angular momentum can be expressed in terms of the radiative multipole moments, which can be then determined in terms of the source multipole moments \cite{Blanchet:1985sp,Blanchet:1987wq,Blanchet:1989ki,Damour:1990ji,Blanchet:1998in,Poujade:2001ie}. 
The leading-order spin-orbit corrections to the needed mass and current multipole moments have been computed by Kidder himself \cite{Kidder:1995zr}.
The resulting expressions for the fluxes  $\frac{dX}{dt}={\mathcal F}_X$ are then given by
\bea
{\mathcal F}_E&=&{\mathcal F}_E^{\rm N}+\eta^2{\mathcal F}_E^{\rm 1PN}+\eta^3{\mathcal F}_E^{\rm SO}
\,,\nonumber\\ 
{\mathcal F}_J&=&{\mathcal F}_J^{\rm N}+\eta^2{\mathcal F}_J^{\rm 1PN}+\eta^3{\mathcal F}_J^{\rm SO}
\,,\nonumber\\ 
{\mathcal F}_{P_i}&=&{\mathcal F}_{P_i}^{\rm N}+\eta{\mathcal F}_{P_i}^{\rm SO}
\,.
\eea

We list below only the spin-orbit corrections to the fluxes, which are of fractional 1.5PN order for energy and angular momentum, and 0.5PN order for linear momentum (see Eqs. (3.25), (3.28), and (3.31) of Ref. \cite{Kidder:1995zr})
\bea
{\mathcal F}_E^{\rm SO}&=&-\frac{8}{15}\frac{M^5}{r^4}\nu^2\dot \phi\left[
v^2(37S+43S_*)-\dot r^2(27S+51S_*)\right.\nonumber\\
&+&\left.
\frac{4M}{r}(3S-S_*)
\right]
\,,\nonumber\\
{\mathcal F}_J^{\rm SO}&=&-\frac{8}{15}\frac{M^4}{r^3}\nu^2\left[
v^4(9S+11S_*)-3\dot r^2v^2(3S+8S_*)\right.\nonumber\\
&+&
\frac{M}{r}(25S+11S_*)v^2+15S_*\dot r^4-\frac{M}{r}(25S+13S_*)\dot r^2\nonumber\\
&+&\left.
\frac{M^2}{r^2}(3S-S_*)
\right]
\,,\nonumber\\
{\mathcal F}_{P_i}^{\rm SO}&=& -\frac{8}{15}\frac{M^3}{r^5}\nu^2\epsilon_{ijz}\left[
-2v^2n^j+4\dot r v^j
\right]
\frac{M}{\delta m}(S-S_*)\,,
\eea
with 
\beq
\frac{ S-S_* }{\delta m}=\frac{S_1}{m_1}-\frac{S_2}{m_2}\,.
\eeq

Integrating the fluxes over time along the hyperboliclike orbit \eqref{orbit} then gives the total energy, angular momentum and linear momentum radiated in gravitational waves during the scattering process
\beq
\Delta X=\int_{-\infty}^\infty dt\,{\mathcal F}_X(t)\,.
\eeq
At the leading PN order the spin-orbit corrections to the orbital averages of the radiative losses turn out to be 
\begin{widetext}
\bea
\label{DeltaE_spin}
(\Delta E)_{\rm SO}&=&-\frac{\nu^2}{[\bar a_r(e_r^2-1)]^5}\left[
(A^{E}_S S+A^{E}_{S_*} S_*){\rm arcos}\left(-\frac{1}{e_r}\right)
+\sqrt{e_r^2-1}(B^{E}_S S+B^{E}_{S_*} S_*)\right]\,,
\eea
with
\bea
A^{E}_S&=& \frac{269}{15}e_r^6 + \frac{1388}{5}e_r^4 + \frac{1552}{3}e_r^2 + \frac{1744}{15}\,,\nonumber\\
A^{E}_{S_*}&=& \frac{244}{15}e_r^6 + \frac{1517}{5}e_r^4 + \frac{2612}{5}e_r^2+ \frac{496}{5}\,,\nonumber\\
B^{E}_S&=&\frac{31901}{225}e_r^4 + \frac{119938}{225}e_r^2 + \frac{19072}{75}\,,\nonumber\\
B^{E}_{S_*}&=& \frac{11084}{75}e_r^4 + \frac{127951}{225}e_r^2 + \frac{50582}{225}\,,
\eea
and 
\bea
\label{DeltaJ_spin}
(\Delta J)_{\rm SO}&=&-\frac{\nu^2}{[\bar a_r(e_r^2-1)]^{7/2}}\left[
(A^{J}_S S+A^{J}_{S_*} S_*){\rm arcos}\left(-\frac{1}{e_r}\right)
+\sqrt{e_r^2-1}(B^{J}_S S+B^{J}_{S_*} S_*)\right]\,,
\eea
with
\bea
A^{J}_S&=& \frac{198}{5}e_r^4 + \frac{928}{5}e_r^2 + \frac{1552}{15}\,,\nonumber\\
A^{J}_{S_*}&=&  \frac{554}{15}e_r^4 + \frac{2696}{15}e_r^2+ 80\,,\nonumber\\
B^{J}_S&=&\frac{64}{5}e_r^4 + \frac{5834}{45}e_r^2 + \frac{1676}{9}\,,\nonumber\\
B^{J}_{S_*}&=& \frac{48}{5}e_r^4 + \frac{6098}{45}e_r^2 + \frac{1364}{9}\,,
\eea
and
\bea
\label{DeltaP_spin}
(\Delta P_x)_{\rm SO}&=& 0
\,,\nonumber\\
(\Delta P_y)_{\rm SO}&=&-\frac{4}{15} \frac{\nu^2}{[\bar a_r(e_r^2-1)]^{9/2}}\left[ 
e_r(3 e_r^4+39e_r^2+28){\rm arcos}\left(-\frac{1}{e_r}\right)
+\frac{\sqrt{e_r^2-1}(71 e_r^4+133 e_r^2+6)}{3e_r}\right]\frac{S-S_*}{\delta m}\,.
\eea 
One can equivalently express the orbital elements $\bar a_r$ and $e_r$ in terms of gauge-invariant variables, i.e., the conserved energy and orbital angular momentum parameters $\bar E$ (or even initial momentum at infinity, $p_\infty$) and $L$ through Eq. \eqref{ar_e_r_vs_EL}.
For what concerns comparison with PN-based results, we have checked that the expressions for $(\Delta E)_{\rm SO}(\bar E,L)$ and $(\Delta J)_{\rm SO}(\bar E,L)$ agree with the linear-in-spin terms (at the leading PN order) of the corresponding ones obtained in Ref. \cite{Cho:2021arx}, Section 5.4, by applying the boundary-to-bound map to existing results for ellipticlike orbits (see also Refs. \cite{Kalin:2019rwq,Kalin:2019inp}).
$(\Delta P_y)_{\rm SO}$ instead is computed here for the first time.

In order to compare with available PM-based results the radiative losses should then be written as series expansion in the large angular momentum parameter or equivalently impact parameter by using the relation \cite{Vines:2017hyw,Vines:2018gqi}
\beq
\label{L_vs_b}
L=\frac{b\sqrt{\gamma^2-1}}{Mh}+\frac{h-1}{2}\left[S\left(1-\frac1h\right)+\frac{S_*}{h}\right]\,,
\eeq
where $h=\sqrt{1+2\nu(\gamma-1)}=E/M=1+\nu\bar E\eta^2$, and $\gamma=\sqrt{1+p_\infty^2\eta^2}$.

We finally obtain the following PM-like relations (either in terms of spin variables $(S,S_*)$ or $(S_1,S_2)$)
\bea
\label{deltaESO_b}
(\Delta E)_{\rm SO}&=&-\nu^2\left[
\frac{\pi p_\infty^2}{b^4}\left(\frac{13}{2}    S+\frac{69}{10}    S_*\right)
+\frac{1}{b^5}\left(\frac{4864 }{25}  S+\frac{40832}{225}   S_*\right)
+\frac{\pi}{ p_\infty^2b^6}\left(\frac{537}{2} S+\frac{455}{2}    S_*\right)
+O\left(\frac{1}{b^7}\right)\right]\nonumber\\
&=&-\nu^2\left\{
\frac{\pi p_\infty^2}{b^4}\left[\left(\frac{13}{2} + \frac{69}{10}\frac{m_2}{m_1}\right)S_1 
+\left(\frac{13}{2} + \frac{69}{10}\frac{m_1}{m_2}\right)S_2\right]\right.\nonumber\\
&+&
\frac{1}{b^5}\left[
\left(\frac{4864}{25} + \frac{40832}{225}\frac{m_2}{m_1}\right)S_1  
+\left(\frac{4864}{25} + \frac{40832}{225}\frac{m_1}{m_2}\right)S_2
\right]\nonumber\\
&+&\left.
\frac{\pi }{p_\infty^2b^6}\left[\left(\frac{537}{2} + \frac{455}{2}\frac{m_2}{m_1}\right)S_1 
+\left(\frac{537}{2} + \frac{455}{2}\frac{m_1}{m_2}\right)S_2\right]
+O\left(\frac{1}{b^7}\right)\right\}
\,,
\eea
and
\bea
\label{deltaJSO_b}
(\Delta J)_{\rm SO}&=&-\nu^2\left[ \frac{32}{5} (S + S_*)  \frac{p_\infty^3  }{b^2}
+\frac{99}{5} \left(S +  \frac{277}{297} S_* \right) \frac{\pi p_\infty }{b^3}
+\frac{12512}{45}\left(S +  \frac{328}{391} S_* \right)\frac{1}{p_\infty b^4}
+O\left(\frac{1}{b^4}\right)\right]\nonumber\\
&=&-\nu^2\left\{
\frac{p_\infty^3}{b^2}\left[
\frac{32}{5}\left(1+\frac{m_2}{m_1}\right)S_1
+\frac{32}{5}\left(1+\frac{m_1}{m_2}\right)S_2
\right]\right.\nonumber\\
&+&
\frac{\pi p_\infty }{b^3}\left[
\left(\frac{99}{5}+\frac{277}{15}\frac{m_2}{m_1}\right)S_1
+ \left(\frac{99}{5}+\frac{277}{15}\frac{m_1}{m_2}\right)S_2
\right]\nonumber\\
&+&\left.
\frac{1}{p_\infty b^4}\left[
\left(\frac{12512}{45}+\frac{10496}{45}\frac{m_2}{m_1}\right)S_1
+\left(\frac{12512}{45}+\frac{10496}{45}\frac{m_1}{m_2}\right)S_2
\right]
+O\left(\frac{1}{b^5}\right)\right\}
\,,
\eea
and 
\bea
\label{deltaPySO_b}
(\Delta P_y)_{\rm SO}&=&-\frac{\nu^2(S-S_*)}{(m_1-m_2)}\left[\frac{2p_\infty\pi}{5b^4}+\frac{64}{9p_\infty b^5}
+\frac{31\pi}{5p_\infty^3b^6}
+O\left(\frac{1}{b^7}\right)\right]\nonumber\\
&=&-\nu^2\left(\frac{S_1}{m_1}-\frac{S_2}{m_2}\right)\left[\frac{2p_\infty\pi}{5b^4}+\frac{64}{9p_\infty b^5}
+\frac{31\pi}{5p_\infty^3b^6}
+O\left(\frac{1}{b^7}\right)\right] 
\,.
\eea
\end{widetext}
for the spin-orbit corrections to the radiated energy, angular momentum and linear momentum, respectively, at the corresponding leading PN order.
Recalling the nonspinning results \cite{Bini:2021gat,Bini:2021jmj,Bini:2021qvf,Bini:2022yrk,Bini:2022xpp}, one sees that at each PM level of accuracy (i.e., each power of $b$) when nonspinning terms contain a $\pi$ the spinning ones do not and viceversa.

\section{Check with existing PM results}

The total radiated energy, angular momentum and linear momentum from spinning binaries have been recently computed in Refs. \cite{Jakobsen:2021lvp,Riva:2022fru} in the framework of the worldline quantum field theory at the leading PM order.
Therefore, these results are valid for arbitrary values of the velocity, but limited to the leading order in $G$.
We can then compare our results with the leading order term of the low-velocity limit of those expressions.
In addition, we can provide higher-order PM terms which can be used for future crosschecks.   

The radiated energy was first computed in Ref. \cite{Jakobsen:2021lvp} Eq. (31), as a low-velocity expansion.
Ref. \cite{Riva:2022fru} then generalized this result by obtaining for it an exact expression valid for all orders in the velocity (see also Ref. \cite{Jakobsen:2022zsx} for an independent computation, with results in complete agreement with \cite{Riva:2022fru}). 
Ref. \cite{Riva:2022fru} also calculated the radiated four momentum at the leading PM level.
However, no explicit expressions were given there, even if the associated Supplemental Material contains all necessary information.
Therefore, we include below a short derivation useful for the comparison.

In the spinless case the four momentum at the 3PM level has the following simple form
\beq
\Delta P^\mu =\pi \left(\frac{G M}{b}\right)^3 M\nu^2  \frac{\hat {\mathcal E}(\gamma)}{\gamma+1}\, (u_1^\mu +u_2^\mu)\,,
\eeq 
stating that $\Delta P^\mu$ is aligned with the (timelike) center-of-velocities vector of $u_1$ and $u_2$, $U_{\rm cv}$ ($U_{\rm cv}\cdot U_{\rm cv}=-1$), namely
\beq
\Delta P^\mu= E(U_{\rm cv})U_{\rm cv}^\mu\,,
\eeq
where
\beq
U_{\rm cv}^\mu=\frac{u_1^\mu +u_2^\mu}{\sqrt{2(1+\gamma)}}\,,
\eeq
and
\beq
E(U_{\rm cv})=\pi \left(\frac{G M}{b}\right)^3 M\nu^2\frac{\sqrt{2}\hat{\mathcal E}(\gamma)}{\sqrt{\gamma+1}} \,.
\eeq
The energy function $\hat{\mathcal E}(\gamma)$ is exactly known as a function of $\gamma=(1-v^2)^{-1/2}$ (see Eq. (7.39) of Ref.  \cite{Herrmann:2021tct}), and has the following low-velocity expansion 
\beq
\hat{\mathcal E}(\gamma)=\frac{37}{15} v+\frac{2393}{840}v^3+\frac{61703}{10080}v^5+O(v^7)\,.
\eeq
For the state of the art for the PM scattering of spinless bodies including radiation-reaction effects see, e.g., Refs. \cite{Bini:2022enm,Dlapa:2022lmu} and references therein.

The radiated four-momentum for spinning bodies at the lowest PM order up to quadratic order in the (arbitrarily oriented) spins has the following general form (see Eq. (18) of Ref. \cite{Riva:2022fru})
\beq
\label{form_gen}
\Delta P^\mu =-\pi \left(\frac{G M}{b}\right)^3 M\nu^2   (C_{u_1}\check u_1^\mu +C_{u_2}\check u_2^\mu+C_{\hat b} \hat b^\mu +C_{\hat l} \hat l^\mu)\,,
\eeq 
where\footnote{
Our notation for $\check u_A$ differs by a sign with respect to that of Ref. \cite{Riva:2022fru}. 
This implies an extra minus sign in front of $C_{u_A}$ in Eq. \eqref{form_gen}.
} 
\bea
\check u_1&=&\frac{u_1-\gamma u_2}{\gamma^2-1}=\frac{P(u_2)u_1}{\gamma^2-1}\,,\nonumber\\
\check u_2&=&\frac{u_2-\gamma u_1}{\gamma^2-1}=\frac{P(u_1)u_1}{\gamma^2-1}\,,
\eea
with $P(u_A)=I+u_A\otimes u_A$ the projector orthogonal to the timelike direction $u_A$, and $\hat b$ and $\hat l$ denote the
unit vectors aligned with the directions of the impact parameter and orbital angular momentum, respectively.
In the case of spins aligned with the orbital angular momentum we are considering here the system radiates momentum only along the direction of the relative velocity of the two bodies, just as in the nonspinning case.
The above expression \eqref{form_gen} thus reduces to
\bea
\label{form_new}
\Delta P^\mu &=&-\pi \left(\frac{G M}{b}\right)^3 M\nu^2   (C_{u_1}\check u_1^\mu +C_{u_2}\check u_2^\mu)\,,\nonumber\\
&=&\pi \left(\frac{G M}{b}\right)^3 M\nu^2   (\widehat C_{u_1} u_1^\mu +\widehat C_{u_2} u_2^\mu)\,,
\eea 
with (see Table I of Ref. \cite{Riva:2022fru})
\bea
C_{u_1}&=& f_{\rm I}(\gamma)+\frac{1}{b}[f_{\rm II}(\gamma)S_1+f_{\rm III}(\gamma)S_2]\,,\nonumber\\
\widehat C_{u_1} &=& -\frac{C_{u_1}-\gamma C_{u_2}}{\gamma^2-1}\,.
\eea
These coefficients are dimensionless functions of $\gamma=-u_1\cdot u_2$ and the spins $S_1$ and $S_2$, and are symmetric under the interchange $1\leftrightarrow 2$, i.e., $C_{u_1}=C_{u_1}(\gamma, S_1,S_2)$ and $\widehat C_{u_1}=\widehat C_{u_1}(\gamma, S_1,S_2)$, with $C_{u_2}=C_{u_1}|_{1\leftrightarrow 2}$ and $\widehat C_{u_2}=\widehat C_{u_1}|_{1\leftrightarrow 2}$.
In absence of spin  $\widehat C_{u_1}=\hat {\mathcal E}(\gamma)/(\gamma+1)$, so that $f_{\rm I}(\gamma)=\hat {\mathcal E}(\gamma)$.
In the PN expansion limit $\gamma=\sqrt{1+p_\infty^2}=1 + \frac12 p_\infty^2 - \frac18 p_\infty^4+O(p_\infty^6)$, and
\bea
f_{\rm II}(\gamma)&=& \frac{13}{2} p_\infty^2 + \frac{1951}{448}p_\infty^4  + \frac{9413}{2688}p_\infty^6 + O(p_\infty^8)\,,\nonumber\\
f_{\rm III}(\gamma)&=& \frac{69}{10} p_\infty^2 + \frac{8917}{2240}p_\infty^4  + \frac{8761}{2688}p_\infty^6  + O(p_\infty^8)\,.
\eea

Let us re-express $u_1$ and $u_2$ in the incoming c.m. frame defined by $U=(m_1u_1+m_2u_2)/E$, $E=E_1+E_2=Mh=M\sqrt{1+2\nu(\gamma-1)}$
and the spatial vectors $\hat {\mathbf b}$, ${\mathbf n}=\frac{1}{p_\infty}\left(\frac{E_2}{m_2}u_1-\frac{E_1}{m_1}u_2 \right)$ (in the orbital plane) and ${\mathbf e}_z$, with $\hat {\mathbf b}$ and ${\mathbf n}$ related to the harmonic-gauge $x$ and $y$ axes adapted to the quasi-Keplerian representation of the orbit by Eqs. (3.49) of Ref. \cite{Bini:2021gat}   as
\bea
\hat {\mathbf b}&=&\cos \frac{\chi_{\rm cons}}{2}{\mathbf e}_x-\sin \frac{\chi_{\rm cons}}{2}{\mathbf e}_y\,,\nonumber\\ 
     {\mathbf n}&=&\sin \frac{\chi_{\rm cons}}{2}{\mathbf e}_x+\cos \frac{\chi_{\rm cons}}{2}{\mathbf e}_y\,,
\eea
such that
\bea
u_1&=&\frac{E_1}{m_1}U +\frac{P_{\rm cm}}{m_1}{\mathbf n}\,,\nonumber\\
u_2&=&\frac{E_2}{m_2}U -\frac{P_{\rm cm}}{m_2}{\mathbf n}\,,
\eea
with $E_a=\sqrt{m_a^2+P_{\rm cm}^2}$, $P_{\rm cm}=m_1m_2 p_\infty/E$.
The four momentum \eqref{form_new} thus becomes
\bea
\Delta P^\mu 
&=&\pi \left(\frac{G M}{b}\right)^3 M\nu^2  \left[ \left(\widehat C_{u_1}  \frac{E_1}{m_1} + \widehat C_{u_2}  \frac{E_2}{m_2} \right) U^\mu\right.\nonumber\\
&+&\left.
P_{\rm cm}
\left(  \frac{ \widehat C_{u_1}}{m_1} -  \frac{ \widehat C_{u_2}}{m_2} \right){\mathbf n}^\mu  \right]\,,
\eea 
so that  
\bea
\Delta P\cdot U&=&-\pi  \frac{G^3  }{b^3}  m_1^2m_2^2   \left(\widehat C_{u_1}  \frac{E_1}{m_1} + \widehat C_{u_2}  \frac{E_2}{m_2} \right)\nonumber\\
&=&-\Delta E
\,,\nonumber\\
\Delta P\cdot {\mathbf n}&=&\pi  \frac{G^3 m_1^3m_2^3 p_\infty}{b^3 E}
\left(  \frac{ \widehat C_{u_1}}{m_1} -  \frac{ \widehat C_{u_2}}{m_2} \right) \nonumber\\
&=&\sin \frac{\chi_{\rm cons}}{2}\Delta P_x+\cos \frac{\chi_{\rm cons}}{2}\Delta P_y\,.
\eea
To linear order in spin and at the leading PM level the spin-dependent terms of the previous relations turn out to be
\bea
(\Delta E)_{\rm SO}&=&\pi  \frac{G^3  }{b^3}  m_1^2m_2^2   \left(\widehat C_{u_1}^{\rm SO}  \frac{E_1}{m_1} + \widehat C_{u_2}^{\rm SO}  \frac{E_2}{m_2} \right)
+O\left(\frac1{b^5}\right)
\,,\nonumber\\
(\Delta P_y)_{\rm SO}&=&\pi  \frac{G^3 m_1^3m_2^3 p_\infty}{b^3 E}
\left(  \frac{ \widehat C_{u_1}^{\rm SO}}{m_1} -  \frac{ \widehat C_{u_2}^{\rm SO}}{m_2} \right) 
+O\left(\frac1{b^5}\right)
\,,
\eea
where (using the coefficients $C_{u_A}^{s_1}$ and $C_{u_A}^{s_2}$ given in the Supplemental Material of Ref. \cite{Riva:2022fru})
\begin{widetext}
\bea
\widehat C_{u_1}^{\rm SO}&=&\widehat C_{u_1}^{s_1}\frac{S_1}{m_1}+\widehat C_{u_1}^{s_2}\frac{S_2}{m_2}
=\frac{1}{b}\left[\left(\frac{2}{5} + \frac{689}{224} p_\infty^2 \right) \frac{S_1}{m_1}  + \left(-\frac{2}{5} + \frac{4059}{1120} p_\infty^2\right) \frac{S_2}{m_2} +O(p_\infty^4)\right]
\,,\nonumber\\
\widehat C_{u_2}^{\rm SO}&=&\widehat C_{u_2}^{s_1}\frac{S_1}{m_1}+\widehat C_{u_2}^{s_2}\frac{S_2}{m_2}
=\frac{1}{b}\left[ \left(-\frac{2}{5} + \frac{4059}{1120} p_\infty^2\right) \frac{S_1}{m_1}  +\left(\frac{2}{5} + \frac{689}{224} p_\infty^2 \right) \frac{S_2}{m_2} +O(p_\infty^4)\right]\,.
\eea
The leading terms in the PN expansion of the energy and linear momentum losses are then given by
\bea
(\Delta E)_{\rm SO}&=&\nu^2\frac{\pi}{b^4}\left\{
\left[\left(\frac{13}{2} + \frac{69}{10}\frac{m_2}{m_1}\right)S_1 
+\left(\frac{13}{2} + \frac{69}{10}\frac{m_1}{m_2}\right)S_2\right] p_\infty^2\right.\nonumber\\
&&
+\left[\left(\frac{1951}{448}-\frac{13}{4}\nu + \left(\frac{8917}{2240}-\frac{69}{20}\nu\right)\frac{m_2}{m_1}\right)S_1 
+\left(\frac{1951}{448}+\frac{13}{4}\nu + \left(\frac{8917}{2240}-\frac{69}{20}\nu\right)\frac{m_1}{m_2}\right)S_2\right]p_\infty^4\nonumber\\
&&
+\left[\left(\frac{39}{16}\nu^2 - \frac{1223}{896}\nu+ \frac{9413}{2688} + \left(\frac{207}{80}\nu^2 - \frac{5053}{4480}\nu + \frac{8761}{2688}\right)\frac{m_2}{m_1}\right)S_1 \right.\nonumber\\
&&\left.\left.
+\left(\frac{39}{16}\nu^2 - \frac{1223}{896}\nu+ \frac{9413}{2688} + \left(\frac{207}{80}\nu^2 - \frac{5053}{4480}\nu+ \frac{8761}{2688}\right)\frac{m_1}{m_2}\right)S_2\right]p_\infty^6
+O( p_\infty^8)\right\}
+O\left(\frac1{b^5}\right)
\,,
\eea
and
\bea
(\Delta P_y)_{\rm SO}&=&\nu^2\frac{\pi}{b^4}\left[
\frac{2}{5}\left(\frac{S_1}{m_1}-\frac{S_2}{m_2}\right) p_\infty\right.\nonumber\\
&&
+\left[\left(-\frac{4059}{1120}-\frac15\nu+\left(\frac{689}{224}-\frac15\nu\right)\frac{m_2}{m_1}\right)S_1
-\left(-\frac{4059}{1120}-\frac15\nu+\left(\frac{689}{224}-\frac15\nu\right)\frac{m_1}{m_2}\right)S_2
\right]p_\infty^3\nonumber\\
&&
+\left[\left(\frac{3}{20}\nu^2 + \frac{4171}{2240}\nu - \frac{4321}{2688}+\left(\frac{3}{20}\nu^2 - \frac{3333}{2240}\nu + \frac{11899}{13440}\right)\frac{m_2}{m_1}\right)S_1\right.\nonumber\\
&&\left.\left.
-\left(\frac{3}{20}\nu^2 + \frac{4171}{2240}\nu - \frac{4321}{2688}+\left(\frac{3}{20}\nu^2 - \frac{3333}{2240}\nu + \frac{11899}{13440}\right)\frac{m_1}{m_2}\right)S_2
\right]p_\infty^5
+O( p_\infty^7)\right]
+O\left(\frac1{b^5}\right)
\,,
\eea
respectively, which agree with our PN-based results, Eqs. \eqref{deltaESO_b} and \eqref{deltaPySO_b}, at the leading PM-PN level, provided taking into account different conventions for the spins.

Finally, the angular momentum has been computed in Ref. \cite{Jakobsen:2021lvp}, Eq. (29).
To linear order in spin the PN expansion of the spin-dependent part reads
\beq
(\Delta J)_{\rm SO}=\frac{\nu^2}{b^2}\left[
\frac{32}{5}p_\infty^3
+\left(-\frac{16}{5}\nu + \frac{16}{35}\right)p_\infty^5
+\left(\frac{12}{5}\nu^2 + \frac47\nu - \frac{172}{315}\right)p_\infty^7 
+O( p_\infty^9)\right]
\left(\frac{S_1}{m_1}+\frac{S_2}{m_2}\right)
+O\left(\frac1{b^3}\right)
\,,
\eeq
\end{widetext}
in the aligned-spin case, and agrees with the leading order term of our result, Eq. \eqref{deltaJSO_b}.

\section{Benchmarks for future PM computations}

Under the approximations done in the present work, namely leading spin-orbit interaction between spinning bodies with constant-in-magnitude spin vectors aligned with the orbital angular momentum, we find it convenient to summarize our results in order to provide ready-to-use expressions for consistency checks with future PM-based computations. 
The radiative losses can be written as a combined PM-PN expansion series, which at the leading PN order reads
\begin{widetext}
\bea
\label{various_expansions}
(\Delta E)_{\rm SO}&=&-\nu^2\sum_{n=4}^{\infty}\frac{1}{b^n}p_\infty^{-2n+10}
\left[E_{n}^{\rm SO}\left(\frac{m_2}{m_1}\right)S_1+E_{n}^{\rm SO}\left(\frac{m_1}{m_2}\right)S_2\right]
\,,\nonumber\\
(\Delta J)_{\rm SO}&=&-\nu^2\sum_{n=2}^{\infty}\frac{1}{b^n}p_\infty^{-2n+7}
\left[J_{n}^{\rm SO}\left(\frac{m_2}{m_1}\right)S_1+J_{n}^{\rm SO}\left(\frac{m_1}{m_2}\right)S_2\right]
\,,\nonumber\\
(\Delta P_y)_{\rm SO}&=&-\nu^2\sum_{n=4}^{\infty}\frac{1}{b^n}p_\infty^{-2n+9}
\left[P_{y\,n}^{\rm SO}\left(\frac{m_2}{m_1}\right)S_1-P_{y\,n}^{\rm SO}\left(\frac{m_1}{m_2}\right)S_2\right]
\,,
\eea
\end{widetext}
where  $E_{n}^{\rm SO}$, $J_{n}^{\rm SO}$ and $P_{y\,n}^{\rm SO}$ are linear functions of the ratio $m_1/m_2$ (and $m_2/m_1$ when exchanging $1\leftrightarrow 2$).
The first few coefficients are listed in Table \ref{tab:EnJnPyn}.


\begin{table}[h]  
\caption{\label{tab:EnJnPyn} 
Coefficients $E_{n}^{\rm SO}$, $J_{n}^{\rm SO}$ and $P_{y\,n}^{\rm SO}$ entering PM expansion \eqref{various_expansions} of the radiated energy, angular momentum, and $y$-component of the linear momentum, respectively, are listed below in functional form at the leading PN order.
At the lowes PM level (i.e., $E_4^{\rm SO}$, $J_{2}^{\rm SO}$ and $P_{y\,4}^{\rm SO}$) the coefficients are exactly known in PN sense.
}
\begin{ruledtabular}
\begin{tabular}{ll}
$E_4^{\rm SO}(x)$ &$\pi\left(\frac{13}{2} + \frac{69}{10}x\right)$\\
$E_5^{\rm SO}(x)$ &$\frac{4864}{25} + \frac{40832}{225}x$\\
$E_6^{\rm SO}(x)$ &$\pi\left(\frac{537}{2} + \frac{455}{2}x\right)$\\
\hline
$J_{2}^{\rm SO}(x)$ &$\frac{32}{5}\left(1+x\right)$\\
$J_{3}^{\rm SO}(x)$& $\pi\left(\frac{99}{5}+\frac{277}{15}x\right)$\\
$J_{4}^{\rm SO}(x)$ &$\frac{12512}{45}+\frac{10496}{45}x$\\
\hline
$P_{y\,4}^{\rm SO}(x)$ &$\frac25\pi(1+x)$\\
$P_{y\,5}^{\rm SO}(x)$ &$\frac{64}{9}(1+x)$\\
$P_{y\,6}^{\rm SO}(x)$ &$\frac{31}{5}\pi(1+x)$\\
\end{tabular}
\end{ruledtabular}
\end{table}

\section{Concluding remarks}

We have studied the gravitational radiation emitted from a spinning binary system undergoing a scattering process due to the leading-order spin-orbit interaction, in the special case of constant spins aligned with the orbital angular momentum, and at the leading PN approximation level.
We have used a leading-PN order spin-modified quasi-Keplerian parametrization of the hyperboliclike orbit, which in the case of constant and aligned spins results in a modification of the orbital elements by linear-in-spin corrections. 
We have then computed the energy, angular momentum and linear momentum losses by the system at their LO in the spin-orbit coupling within the PN-MPM approach to gravitational radiation by assuming the balance between radiative losses in the
near zone of the source and radiation fluxes in the far zone. 

We found agreement with existing PN-based results \cite{Cho:2021arx}, where the radiated energy and angular momentum were obtained by using the boundary-to-bound correspondence with ellipticlike orbits.
The leading PN expression for the linear momentum loss is instead given here for the first time.
For what concerns PM-based results, we have confirmed recent accomplishments obtained in Refs. \cite{Jakobsen:2021lvp,Riva:2022fru} by using a worldline quantum field approach for the radiated energy, angular momentum and linear momentum a the lowest PM level.
We have also provided higher-order PM corrections as possible benchmarks for future computations nowadays at hands.
Finally, extensions of these results including, e.g., higher-order terms in the leading-order PN expansion, and next-to-leading order terms in the spin-orbit and higher-order spin-spin couplings are straightforward, and will be considered in forthcoming works.

\section*{Acknowledgments}
The authors thank R. Porto for useful discussions.
DB and PR are grateful to IHES for the hospitality and support (by the \lq\lq 2021 Balzan  
Prize for Gravitation: Physical and Astrophysical  
Aspects," awarded to T. Damour) in such a  highly stimulating environment during the development of the present project. 
P.~R. is supported by the Italian Minister of University and Research (MUR) via the 
PRIN 2020KB33TP, {\it Multimessenger astronomy in the Einstein Telescope Era (METE)}.
D.B.  acknowledges sponsorship of the Italian Gruppo Nazionale per la Fisica  
Matematica (GNFM) of the Istituto Nazionale di Alta  
Matematica (INDAM).

\end{document}